\documentclass[conference]{IEEEtran}
\IEEEoverridecommandlockouts
\usepackage{cite}
\usepackage{amsmath,amssymb,amsfonts,amsthm}
\usepackage{algorithmicx}%
\usepackage{algpseudocode}%
\usepackage{listings}%
\usepackage{algorithm}
\usepackage{graphicx}
\usepackage{textcomp}
\usepackage{xcolor}
\def\BibTeX{{\rm B\kern-.05em{\sc i\kern-.025em b}\kern-.08em
    T\kern-.1667em\lower.7ex\hbox{E}\kern-.125emX}}
\begin{document}

\renewcommand{\algorithmicrequire}{\textbf{Parameter:}}
\renewcommand{\algorithmicensure}{\textbf{Output:}}
\algnewcommand\algorithmicforeach{\textbf{For Each:}}
\algnewcommand\ForEach{\item[ \algorithmicforeach]}
\algnewcommand\algorithmicqueryphase{\textbf{Query Phase:}}
\algnewcommand\QueryPhase{\item[ \algorithmicqueryphase]}
\algnewcommand\algorithmictrainingphase{\textbf{Training Phase:}}
\algnewcommand\TrainingPhase{\item[ \algorithmictrainingphase]}
\algnewcommand\algorithmicallocationphase{\textbf{Allocation Phase:}}
\algnewcommand\AllocationPhase{\item[ \algorithmicallocationphase]}
\algnewcommand\algorithmicpaymentphase{\textbf{Payment Calculation:}}
\algnewcommand\PaymentPhase{\item[ \algorithmicpaymentphase]}

\newtheoremstyle{bfnote}%
  {}{}
  {\itshape}{}
  {\bfseries}{.}
  { }{\thmname{#1}\thmnumber{ #2}\thmnote{ (#3)}}

\theoremstyle{bfnote}%
\newtheorem{theorem}{Theorem}
\newtheorem{definition}{Definition}
\newtheorem{lemma}{Lemma}

\title{Differentially Private Machine Learning-powered Combinatorial Auction Design\\
\thanks{\footnotesize \textsuperscript{*}These authors equally contributed in this research.}
}

\author{\IEEEauthorblockN{Arash Jamshidi\textsuperscript{*}}
\IEEEauthorblockA{\textit{Department of Computer Engineering} \\
\textit{Sharif University of Technology}\\
arashjamshidi@sharif.edu}
\and
\IEEEauthorblockN{Seyed Mohammad Hosseini\textsuperscript{*}}
\IEEEauthorblockA{\textit{Department of Computer Engineering} \\
\textit{Sharif University of Technology}\\
semo.hosseini@sharif.edu}
\and
\IEEEauthorblockN{Seyed Mahdi Noormousavi}
\IEEEauthorblockA{\textit{Department of Computer Engineering} \\
\textit{Sharif University of Technology}\\
mahdi.noormousavi75@sharif.edu}
\and
\IEEEauthorblockN{Mahdi Jafari Siavoshani}
\IEEEauthorblockA{\textit{Department of Computer Engineering} \\
\textit{Sharif University of Technology}\\
mjafari@sharif.edu}

}

\maketitle

\begin{abstract}
We present a new approach to machine learning-powered combinatorial auctions, which is based on the principles of Differential Privacy. Our methodology guarantees that the auction mechanism is truthful, meaning that rational bidders have the incentive to reveal their true valuation functions. We achieve this by inducing truthfulness in the auction dynamics, ensuring that bidders consistently provide accurate information about their valuation functions.

Our method not only ensures truthfulness but also preserves the efficiency of the original auction. This means that if the initial auction outputs an allocation with high social welfare, our modified truthful version of the auction will also achieve high social welfare. We use techniques from Differential Privacy, such as the Exponential Mechanism, to achieve these results. Additionally, we examine the application of differential privacy in auctions across both asymptotic and non-asymptotic regimes.


\end{abstract}


\section{Introduction}
In recent years, there has been a noticeable increase in the development of innovative machine learning methods for designing Combinatorial Auctions (CA). This is evident in the works of Brero et al. (2021) and Lubin et al. (2021) \cite{brero_machine_2021,beyeler2021imlca}. These approaches involve collecting data from bidders by asking them about the value of certain bundles. The collected data is then used to develop machine learning models that estimate the bidders' valuation functions.

In practical scenarios where bidders truthfully report their valuation for each bundle in their query set, these machine learning-based approaches prove to be effective. This results in high social welfare in the final allocation \cite{brero_machine_2021}. However, when bidders deviate from truthfulness and engage in strategic behaviors to maximize their utility, predicting the auction's outcome and achieving optimal efficiency become a challenging problem.

Differential Privacy methods and concepts are becoming increasingly popular in various practical and theoretical scenarios. Differential privacy has primarily been used to ensure the confidentiality of sensitive information and protect the privacy of individuals. However, recent research has shown that differential privacy can be used for other purposes as well. In the context of Mechanism Design, new approaches have emerged that utilize differential privacy not only to ensure privacy but also to ensure other useful attributes such as Truthfulness.


The aim of this paper is to propose a Differential Privacy method that can modify any machine learning-based auction to encourage bidders to provide truthful information consistently. The proposed method has been developed with the objective of improving auction efficiency by mitigating the impact of strategic behaviors. The research establishes that, under certain assumptions, the final efficiency of the differentially private auction closely approximates that of the original auction where all bidders were truthful. We have analyzed our method in both asymptotic and non-asymptotic regimes concerning the number of bidders, which we denote as $n$.

\subsection{Related Works}
In recent years, researchers have explored the intersection of Differential Privacy, Mechanism Design, and Game Theory. This interdisciplinary field has seen a surge in innovative approaches, particularly in the auction mechanisms for selling differentiable private data. Notable contributions include works by Ghosh et al. (2011), Li et al. (2014), and Nissim et al. (2014) \cite{ghosh2011selling,li2014theory,nissim2014redrawing}, which investigate auction mechanisms for selling differentiable private data.

Further research has proposed innovative algorithms for the design of truthful auctions based on differential privacy, such as works by Nissim et al. (2012) and McSherry (2007) \cite{nissim2012approximately, mcsherry2007mechanism}. Another interesting approach involves the application of the Exponential Mechanism to design truthful mechanisms in auctions, as exemplified by the research conducted by Huang et al. (2012) \cite{huang2012exponential}.

Moreover, the intersection of convex optimization, mechanism design, and differential privacy has initiated a new line of research that aims to model and address mechanism design problems by utilizing differential private convex optimization methods. This is evidenced in the work of Hsu et al. (2016) \cite{hsu2016jointly}.

\section{Preliminaries}

\subsection{Combinatorial Auctions}
Below are the foundational aspects of CAs that are necessary for subsequent discussions in the paper.

\subsubsection{Formal Model}
In general, any auction is mainly composed of three components:

\begin{itemize}
	\item The set of potential bidders: $N = \{1, 2, 3, \ldots, n\}$,
	\item The set of all possible resource allocations: $\Omega$,
	\item valuation to each allocation for each bidder $i$: $v^i : \Omega \rightarrow \mathbb{R}^+$.
\end{itemize}

In this paper, we state the formal framework for Combinatorial Auctions (CAs), which involve a group of $n$ participants, referred to as bidders, and a collection of $m$ indivisible items labeled as $M = \{1, 2, \ldots, m\}$. The objective is to distribute the auction items among the bidders in a way that maximizes their social welfare, as defined later.

We denote the set of possible bundles as $\mathcal{X} = \{0,1\}^m$, where each component signifies the presence or absence of a particular item within the bundle. Additionally, each bidder has a valuation function $v^i: \mathcal{X} \rightarrow \mathbb{R}^+$ that captures their private preference. For any bundle $x \in \mathcal{X}$, $v^i(x) \in \mathbb{R}^+$ represents the actual value that bidder $i$ places on acquiring that specific bundle. We assume that for each bidder $i$, $v^i(\emptyset) = 0$ (this is known as the normalization assumption). The collective set of valuation functions, i.e., valuation profile, is denoted as $\mathbf{v} = (v^1, v^2, \ldots, v^n)$.

Under a ``combinatorial auction mechanism,'' the auctioneer and bidders interact to determine the optimal distribution of available items. The participants follow certain rules to allocate the items, which we will define later.
\begin{enumerate}
	\item \textbf{Allocation Rule}:
    To distribute items among bidders, we use an allocation system that takes into account the current bids. Each allocation is represented as $\mathbf{a} = (a^1, a^2, \ldots, a^n) \in \mathcal{X}^n$, where $a^i$ indicates the bundle allocated to the $i$th bidder. It's important to note that each allocation must be feasible, meaning that no item can be assigned to more than one bidder.
	\item \textbf{Payment Rule}:
	After the allocation process, each bidder is expected to make a payment for the bundle they have been assigned. These payments can be represented as a vector $\mathbf{p} = (p^1, p^2, \ldots, p^n) \in \mathbb{R}^n$, where $p^i$ refers to the payment amount assigned to bidder $i$.
\end{enumerate}

In the following, we assume that bidders have a \textit{quasi-linear} utility function \cite{milgrom_putting_2003}, namely,
\begin{equation}
	u^i(\mathbf{a}, \mathbf{p}) = v^i(a^i) - p^i,
\end{equation}
and define \textit{social welfare} of each allocation $\mathbf{a}$ as the sum of the bidders' true values for the bundles, i.e.,
\begin{equation}
	V(\mathbf{a}) = \sum_{i\in N} v^i(a^i).
\end{equation}

Now, we discuss the report of bundle values for bidder $i$. The report is denoted as $(x_{ik},\hat{v}_{ik})$, where $k$ represents the value sequence and $\hat{v}$ represents the estimated value. It is important to note that the values reported during the query phase may not necessarily be truthful or accurate. Therefore, we represent them as $\hat{v}$ instead of $v$. 

In addition, we represent the collection of all bundle-value reports for bidder $i$ as $R_i = {\{(x_{ik},\hat{v}_{ik})\}}_{k\in L}$, where $L = \{1,2, \ldots, l\}$. During the \emph{training phase}, we denote the learned estimated value function through deep neural networks as $\hat{v}^i(\cdot)$.

\subsubsection{Vickrey-Clarke-Groves (VCG) Mechanism}\label{subsec:vcg}
The Vickrey auctions, also known as VCG auctions \cite{vickrey_counterspeculation_1961,clarke_multipart_1971,groves_incentives_1973}, are a unique type of auction where the winning bidder is charged the lowest accepted bid rather than their own higher bid. This approach promotes transparency and eliminates some of the strategic issues that arise in traditional first-price auctions.
Formally, the VCG mechanism is defined as below.

\begin{definition}{(VCG Mechanism \cite{milgrom_putting_2003})}
	The allocation and payment rules in a VCG mechanism are as follows:

\begin{itemize}
	\item \textbf{Allocation Rule}: The allocation rule in this mechanism is the social welfare maximizing allocation, namely,
	\begin{equation}
	\label{eq:alloc}
		o^* \in \arg\max_{o\in \Omega} \sum_i v^i(o)
	\end{equation}
	
	\item \textbf{Payment Rule}: The payment rule is determined by the difference between optimal social welfare before and after participation, i.e.,
	\begin{equation}
	\label{eq:paym}
		p^i = \sum_{j\neq i} v^j(o^{-i}) - \sum_{j\neq i} v^j(o^*),
	\end{equation}
	where $o^{-i} \in \arg\max_{\tilde{o}\in \Omega} \sum_{j \neq i} v^j(\tilde{o})$.
\end{itemize}
\end{definition}
A weakly dominant strategy is the best strategy for a bidder. If the bidder employs another strategy, their utility will be less than or equal to the utility of the weakly dominant strategy.

\begin{definition}{(Truthfulness)}
    The mechanism $\mathcal{M}$ is truthful if each bidder's best strategy is to report their true valuation function.
\end{definition}

In simple terms, a bidder must pay for their participation in an auction. The VCG mechanism is known to be "strategy-proof," which means that each bidder's truthful reporting is the best course of action in this mechanism, according to Milgrom's book (2003) \cite{milgrom_putting_2003}.

The VCG mechanism's fundamental concept is to encourage bidders to provide truthful information, ensuring that the resulting allocation corresponds to the maximum achievable social welfare, as explained in Milgrom's book (2003)\cite{milgrom_putting_2003}.

Under broader circumstances, the VCG mechanism operates by setting market-clearing prices based on marginal externalities. Each individual receives their net social gain, which is calculated as the difference between the total revenue obtained and the combined costs of production and negative externalities generated. If equilibrium criteria are met, individuals responsible for positive externalities are compensated at the same rate as those accountable for negative ones, thereby effectively accounting for these secondary effects. Additionally, these auctions must follow the individual rationality principle, meaning each bidder should have a non-negative utility function after the auction's execution.

We also formally define a special case of the VCG mechanism known as second-price auctions.
\begin{definition}{(Second-price auctions)}
   A second-price auction is a mechanism used to allocate one item to one of the $n$ bidders, using the VCG mechanism for allocation and payment rules. This means that the item is awarded to the bidder with the highest bid and they pay the second-highest bid as payment.
\end{definition}

\subsection{Differential Privacy}
Differential privacy is a data privacy approach that aims to protect individual privacy while allowing for meaningful statistical analysis. It involves adding random noise to datasets to prevent individuals from being identified based on their specific data points. The ultimate goal is to protect sensitive information while preserving the quality of data analysis.

Various methods can be used to achieve differential privacy, including adding random perturbations to datasets, injecting noise into queries performed on datasets, or generating synthetic datasets. As data collection and sharing continue to expand across various fields such as healthcare, finance, and social media, the importance of differential privacy has grown.

In this study, we use differential privacy as a toolkit to introduce controlled modifications into the auction design to ensure its truthfulness. Throughout the rest of this paper, we refer to $\mathcal{R}$ as the collection of all possible datasets. We also introduce a mechanism denoted as $\mathcal{M}: \mathcal{R}^m \rightarrow \mathcal{O}$, which is a potentially randomized mapping from the dataset obtained to the resulting outcomes.

Furthermore, we establish the distance between two datasets, $d^1, d^2 \in \mathcal{R}^m$, as the number of differences in the dataset of corresponding participants, namely,
\begin{equation}
	\text{dist}(d^1,d^2) = |\{i\in [m]: d^1_i \neq d^2_i\}|.
\end{equation}

In the subsequent sections, we will engage with three central interconnected notions:

\begin{definition}{(Differential Privacy \cite{dwork_algorithmic_2013})}
A mechanism $\mathcal{M}: \mathcal{R}^m \rightarrow \mathcal{O}$ is $\epsilon-$differentially private mechanism if for every $d^1, d^2 \in \mathcal{R}^m$ where $\text{dist}(d^1, d^2) \leq 1$, we have,
$	\forall S \subseteq \text{Range}(\mathcal{M}): \quad \mathbb{P}[\mathcal{M}(d^1) \in S] \leq e^{\epsilon} \mathbb{P}[\mathcal{M}(d^2) \in S]$.
\end{definition}

\begin{definition}{($\epsilon-$truthfulness \cite{dwork_algorithmic_2013})}
Given a randomized mechanism $\mathcal{M}: \mathcal{R}^m \rightarrow \Delta(\mathcal{O})$ (e.g. an auction) is $\epsilon$-truthful if for every player $i$ and  every $d_i, d'_i \in \mathcal{R}$ we have,
$    \mathbb{E}[u^i(d_i, \mathcal{M}(d_i, d_{-i}))] \geq \mathbb{E}[u^i(d_i, \mathcal{M}(d'_i, d_{-i}))] - \epsilon$,
where $d_{-i}$ denotes the dataset obtained from players other than $i$, and the $d_i$ could be considered as the type of the bidder in this definition.
\end{definition}


The subsequent crucial idea is formulated to address scenarios where our goal is to select the maximum among some set of outcomes. In this idea, using randomization, we make our mechanism differential private while preserving the quality of our final outcome with high probability.

\begin{definition}{(Exponential Mechanism \cite{dwork_algorithmic_2013})}
Given an $\varepsilon$  parameter and an arbitrary range $\mathcal{T}$ that affects the utility function $w: \mathcal{R}^m \times \mathcal{T} \rightarrow \mathbb{R}$, we define,
	$\Delta w = \max_{t\in \mathcal{T}} \max_{\mathrm{dist}(d^1, d^2)\leq 1} |w(d^1,t) - w(d^2,t)|. $

Then, in the exponential mechanism $\mathcal{M}_E(d,u)$, each output $t \in \mathcal{T}$ in the range has probability proportional to $e^{\frac{\varepsilon w(d,t)}{\Delta ws}}$. We define $\Delta w$ as the sensitivity parameter of utility function $w$. 
\end{definition} 


\begin{theorem}
\label{thm:2eps}
If the range of utilities for all players are in $[0,1]$ and If the mechanism $\mathcal{M}$ is $\epsilon$-differentially private then $\mathcal{M}$ is also $2\epsilon$-truthful.
\end{theorem} 

\textit{Proof.}
Using Proposition 10.1 of \cite{dwork_algorithmic_2013}, the theorem is trivial
\qed

\section{Machine Learning-powered Combinatorial Auctions}
In this section, we will explain how machine learning is utilized in the design of combinatorial auctions. Our framework is based on the work of \cite{brero_machine_2021} and is a simplified version of the MLCA. In the original MLCA, there are multiple rounds of queries during the auction execution. However, in our simplified version, we assume only a single query phase at the initial stage.

During the initial query phase, information is gathered about the value of $T$ randomly selected bundles from each bidder. Bidders have the option to be truthful and report accurate values for each bundle. However, their behavior may not necessarily align with complete truthfulness.

In the subsequent training phase, a \textbf{deterministic} machine learning model is trained separately for each bidder using the corresponding dataset obtained from the query phase. Various loss functions may be employed in this phase. At the end of the training process, we have an estimated function for the valuation of each bidder separately.

Following the training phase, the allocation phase is executed. Based on the acquired knowledge of valuation functions, we try to determine the optimal allocation that maximizes social welfare.

In the final phase, namely the payment calculation, the computation of ultimate payments for each bidder is carried out using the VCG payment rule.

\section{Differentially Private Auction Design}
In this section, we will introduce certain adjustments to the algorithm to ensure truthfulness in scenarios characterized by asymptotic or non-asymptotic number of bidders.

We divide our technical discussion into two domains:
\begin{enumerate}
	\item In the initial domain, we modify our learning algorithm to ensure asymptotic truthfulness. This implies that as the number of bidders, denoted as $n$, approaches infinity, the auction will eventually be truthful.
 
	\item The second domain explores the approach of achieving exact truthfulness in an auction mechanism for bidders. This implies that, under specific assumptions, all rational bidders are compelled to reveal truthful information to maximize their expected utility. A distinctive feature of this method, in contrast to the previous one, is that the necessity for the number of bidders, denoted as $n$, to approach infinity is not a prerequisite for ensuring truthfulness.
\end{enumerate}


\subsection{Truthful Algorithm}

In this section, we simplify the discussion by assuming that all bidders, denoted by $i$, and all bundles, represented by $x$, have valuation functions $v^i(x)$ and estimated valuation functions $\hat{v}^i(x)$ that fall within the range of $[0,1]$.

According to Theorem \ref{thm:2eps}, bidders cannot increase their utility by more than $2\epsilon$ by providing non-truthful value information. This observation can be used to design mechanisms that achieve either approximate or precise truthfulness.


The refined design is illustrated in Algorithm \ref{alg:refdsca}.
\begin{algorithm}
\caption{Refined MLCA}\label{alg:refdsca}
\begin{algorithmic}[1]
\Require $T$: number of queries, machine learning model $\mathcal{X}$
\Ensure Allocation $\hat{a}$ of items and the Payment vector $\hat{p}$
\QueryPhase
\ForEach{Bidder $i \in N$}
\State Generate $T$ random bundles (There are ${2^m}\choose{T}$ of such queries).
\State Ask every bidder the generated queries \label{line:repo}.
\State Make the bundle-value pairs for the reports $R_i$.
\TrainingPhase
\ForEach{Bidder $i \in N$}
\State Train model $\mathcal{X}$ to obtain estimated $\hat{v}^i(\cdot)$, on the reported dataset $R_i$.
\State Stop the training when an affordable loss and number of epochs is obtained.
\AllocationPhase
\State Given the profile of trained neural value function $\hat{v} = (\hat{v}^1, \hat{v}^2, \ldots, \hat{v}^n)$:
\State From the learned valuations $\hat{v}$, find $k$ of the allocations with the highest social welfare with respect to $\hat{v}$’s. (for any of the $n^m$ allocations $a$, calculate social welfare: $sw(a) =\frac{1}{n}\sum_{i=1}^{n}\hat{v}^i(a)$ and return $k$ of them with highest welfares)\label{line:alloc},
in other words, for any k-allocation $A_k=[a_1,a_2,...,a_k]$ (where each is $a_i$ is an allocation) pick $A_k$ to maximize $SW(A_k)=\frac{1}{k}\sum_{a_i\in A_k} sw(a_i)$.
\State 
Given the optimal k-allocation $A_k$, run VCG with $\Omega = A_k$ and output the best allocation. It means that we ask the value of each of these $k$ outputs in each allocation from $A_k$ and then output the one with highest social welfare.
\PaymentPhase
\State Use the VCG payment rule ($\Omega = A_k$) to calculate the payments.
\end{algorithmic}
\end{algorithm}



It is evident that stage \ref{line:alloc} of this algorithm operates deterministically, and rational bidders are incentivized to truthfully report their valuations due to the truthful nature of the VCG mechanism. However, in stage \ref{line:repo}, bidders have the ability to manipulate the auction by providing false valuations. Given the complex nature of our method, which involves the utilization of learning algorithms such as neural networks, effecting changes to induce truthfulness into the auction format proves to be a challenge. Nonetheless, this appears to be potential for addressing this challenge through methods based on differential privacy.

To tackle this issue of truthfulness, let us consider implementing a modification to stage 3 (Allocation Phase) of the algorithm:

\begin{enumerate}
\setcounter{enumi}{6}
	\item  Calculate social welfare of each k-allocation $A_k$ based on learned valuations, but instead of picking the $A_k$ with highest value, pick it using exponential-mechanism with parameter $\Delta=\frac{1}{n}$. In other words, pick from the distribution such that we pick each $A_k$ with probability $\propto \exp (\frac{\epsilon n SW(A_k)}{2})$  and name it $A_{k}^{*}$ (for some $\epsilon \leq1$)
\end{enumerate}

Because $v^i(x)\in[0,1]$ It’s obvious that $sw(x)$ is $\Delta=\frac{1}{n}$-sensitive for all $x$. Meaning bidder $i$ can only change $sw(x)$ by $\frac{1}{n}$ by changing his report. Now in this way our auction is $\epsilon$-differentially private, hence is $2\epsilon$-truthful. Now by the property of exponential-mechanism we have:

\begin{theorem}
Consider we have a Refined MLCA (Algorithm 1) combinatorial auction with the modified changes (using an exponential mechanism and replacing the line 7 of the algorithm as stated before):
\begin{equation*}
SW(A_{k}^{*}) \geq \max_{A_k=[a_{i_{1}},\ldots,a_{i_{k}}]}SW(A_k) - O\left(\frac{mk\log{n}}{\epsilon n}\right),    
\end{equation*}
with high probability.
\label{thm:refined}
\end{theorem}
\textit{Proof.} For the proof see Theorem 3.11 in \cite{dwork_algorithmic_2013}.
\qed

Now if we set $\epsilon$ to something like $\epsilon=O(\frac{1}{\log{n}})$ our auction is asymptotically exactly truthful (as $n$ grows). With large enough $n$ all bidders has no incentive to bid non-truthfully because in the best case scenario they can only increase their utility by at most $O(\frac{1}{\log{n}})$ which is negligible. Hence we have the following theorem.

\begin{theorem}{(Asymptotically Exactly Truthful Auction)}
	If for large enough $n$ we have $m = o(\frac{n}{(\log{n})^{2}})$ then using exponential-mechanism we can design an auction that is  asymptotically exactly truthful and guarantee that the value of final set of $k$ elements $A_{k}^{*}$ is very close to $\max_{k}A_k$. In other words, with high probability:
\begin{equation*}
SW(A_{k}^{*}) \geq \max_{A_k[a_{i_{1}},\ldots,a_{i_{k}}]}SW(A_k) - O\left(\frac{mk(\log{n})^{2}}{n}\right).    
\end{equation*}
\end{theorem} 
\textit{Proof.} By the result of Theorem 2 and by setting $\epsilon = O(\frac{1}{\log{n}})$ and the fact that $m = o(\frac{n}{(\log{n})^{2}})$ we can conclude the theorem.
\qed

We can see that for large enough $n$, the error term $O(\frac{mk(\log{n})^{2}}{n})$ tends to zero.

In this particular scenario, the effectiveness of our learning algorithms becomes important. If these algorithms exhibit a reasonable generalization error, resulting in an allocation with high social welfare, we can reasonably anticipate a comparable level of social welfare with the differentially private algorithm, with high probability.

However, considering scenarios where asymptotic truthfulness may not suffice, and there is a desire for exact truthfulness, the incorporation of differential privacy proves to be particularly valuable. Under certain additional assumptions, we can use differential privacy to ensure not just asymptotic truthfulness but exact truthfulness in the non-asymptotic regime. 

To examine this scenario, we first establish related definitions and assumptions to facilitate our theoretical analysis in the rest of this section.

\begin{definition}
	For any bundle $x$ we define a punishing mechanism $AU^P(x,v)$ as a sealed-bid second-price auction for selling that bundle. $v$ stands for valuation functions of the bidders (or equivalently bids that they place on bundle $x$).
\end{definition} 

Before continuing we make these assumptions:
\begin{enumerate}
	\item \textbf{Assumption 1}: For all bundles $x$ and bidders $i$ we have $v^i(x)\in\{0, c, 2c, \ldots, 1\}$ and bidders can only place bids according to this set meaning $b^i(x)\in\{0, c, 2c, \ldots, 1\}$ for some small $c$.
	\item \textbf{Assumption 2}: Bidders has some valuation function $v$ which shows their value for any bundle. Bidders can lie and report valuations based on some other valuation function $v^{'}$ such that $v^{'} \neq v$.
	\item \textbf{Assumption 3}: Each bidder $i$ believes if he report based on valuation function $v'^{i} \neq v^i$, there is some bundle $x$ such that  $u^i(v^i,AU^P(x,v)) > u^i(v^i,AU^P(x,(v'^{i},v^{-i})))$.
	\item \textbf{Assumption 4}: Number of items is very smaller than number of bidders such that $2^mm=o(\frac{n}{\log{n}})$.

\end{enumerate}

Because everything is multiple of $c$  we can conclude from assumption 3 the following lemma

\begin{lemma}
If Assumption~3 holds, there is some bundle $x$ such that $u^i(v^i,AU^P(x,v)) \geq u^i(v^i,AU^P(x,(v'^{i},v^{-i}))) + c$.
\end{lemma}

Under these assumptions we design our exactly truthful auction $AU^P_{\epsilon} = qAU^P +(1-q)AU_{\epsilon}$ is shown in Algorithm \ref{alg:trdsca}.

\begin{algorithm}
\caption{Truthful MLCA}\label{alg:trdsca}
\begin{algorithmic}[1]
\Require $T$: number of queries, machine learning model $\mathcal{X}$
\Ensure Allocation $\hat{a}$ of items and the Payment vector $\hat{p}$
\QueryPhase
\ForEach{Bidder $i \in N$}
\State Generate $T$ random bundles (There are ${2^m}\choose{T}$ of such queries)
\State For each of the random queries $x$ ask each bidder to report $v^i(x)$.
\State Make the bundle-value pairs for the reports $R_i$.
\State With probability $q$ go to line \ref{line:punish}, else (with probability $1-q$) go to line \ref{line:train}.
\State Pick one of the $T$ bundles randomly and name it $x^*$. Run $AU^P(x^*,v)$ where $v$ is the answered report of line 2 (allocate bundle $x^*$ and charge the winner based on second-price auction rules). return (The auction will end in this stage) \label{line:punish}.
\TrainingPhase
\ForEach{Bidder $i \in N$}
\State Train model $\mathcal{X}$ to obtain $\hat{v}^i(\cdot)$, on the reported dataset $R_i$. \label{line:train}
\State Stop the training when an affordable loss and number of epochs is obtained.
\AllocationPhase ($AU_{\epsilon}$)
\State Given the profile of trained neural value function $\hat{v} = (\hat{v}^1, \hat{v}^2, \ldots, \hat{v}^n)$:
\State Calculate social welfare of each k-allocation $A_k$ based on learned valuations, but instead of picking the $A_k$ with highest value, pick It using exponential-mechanism with parameter $\Delta=\frac{1}{n}$. In other word Pick from the distribution such that we pick each $A_k$ with probability $\propto \exp (\frac{\epsilon n SW(A_k)}{2})$  and name it $A_{k}^{*}$ (for some $\epsilon \leq1$)
\State 
Given the optimal k-allocation $A_k$, run VCG with $\Omega = A_k$ and output the best allocation. It means we ask the value of each of these $k$ outputs in each allocation from $A_k$
\PaymentPhase
\State Use the VCG payment rule ($\Omega = A_k$) to calculate the payments.
\end{algorithmic}
\end{algorithm}

The only thing we added is the randomization between our differential private auction (designated by $AU_{\epsilon}$) and punitive second-price auction $AU^P$. First we’ll see that for a reasonable value of $\epsilon$ we can ensure exact truthfulness. But before that based on assumption 3 it’s obvious that second-price auction $AU^P$ is strictly truthful.

\begin{lemma}
Based on Lemma 1 and the fact that second-price auction is truthful we have:
\begin{flalign}
    \mathbb{E}_{x^*}[&u^i(v^i,AU^P(x^*,v))] \geq \\
&\mathbb{E}_{x^*}[u^i(v^i,AU^P(x^*,(v'^{i},v^{-i})))] + \frac{c}{2^m}.  \nonumber
\end{flalign}
   
\end{lemma} 
\textit{Proof.} 
Based on assumption 3, there is at least one bundle such that if bidder $i$ use valuation function $v' \neq v$, we can find at least one bundle $x$ such that  $u^i(v^i,AU^P(x,v)) > u^i(v^i,AU^P(x,(v'^{i},v^{-i})))$. We also know that in second-price auctions, truthful bidding is the  weakly dominant strategy, so for other bundles, the best possible strategy is the truthful bidding. Based on these and the fact that there are $2^m$ bundles, we can conclude the result. \qed

\begin{theorem}
\label{thm:truthful}
	If Assumptions 1-4 holds, for $2\epsilon \leq \frac{qc}{2^m}$ our auction $AU^P_{\epsilon}$ is truthful.
\end{theorem}
\textit{Proof.} 
\begin{flalign*}
	 u^i(v^i,&AU^P_{\epsilon}(v^i,v^{-i})) \\
	&= (1-q) \cdot u^i\left(v^{i}, AU_{\epsilon}\left(v^{i}, v^{-i}\right)\right) \\
 & \quad \quad + q \cdot u^i\left(v^{i}, AU^{P}\left(v^{i}, v^{-i}\right)\right) \\
	& \geq (1-q)\left(u^i\left(v^{i}, AU_{\epsilon}\left(v^{\prime i}, v^{-i}\right)\right)-2 \epsilon\right) \\
 & \quad \quad +q\left(u^i\left(v^{i}, AU^{P}\left(v^{\prime i}, v^{-i}\right)\right)+\frac{c}{2^m}\right) \\
	&= u^i\left(v^{i}, AU_{\epsilon}^{P}\left(v^{\prime i}, v^{-i}\right)\right)-(1-q) 2 \epsilon+q \frac{c}{2^m} \\
	&= u^i\left(v^{i}, AU_{\epsilon}^{P}\left(v^{\prime i}, v^{-i}\right)\right)-2 \epsilon+q\left(2 \epsilon+\frac{c}{2^m}\right) .
\end{flalign*}

Since we have $2\epsilon \leq \frac{qc}{2^m}$, it completes the proof.
\qed

Using this theorem, we also can bound $SW(A_{k}^{*})$ with high probability as follows:
\begin{theorem}
	According to assumptions of Theorem \ref{thm:truthful} (assumptions 1-4), by setting $\epsilon = O\left(\sqrt{\frac{c m\log{n}}{2^m n}}\right)$ and $q =\frac{2\epsilon 2^m}{c}$, we have:
 \begin{equation*}
    SW(A_{k}^{*}) \geq  \max_{A_k[a_{i_{1}},\ldots,a_{i_{k}}]}SW(A_k) -O\left(\sqrt{\frac{2^mm \log{n}}{c n}}\right), \nonumber
\end{equation*}
with high probability.
\end{theorem}
\textit{Proof.}
\begin{flalign}
	\mathbb{E}_q[SW(A_{k}^{*})] &\geq (1-q) SW_{AU_{\epsilon}}(A_{k}^{*}) \nonumber \\
&\hspace{-20pt} = (1-\frac{2\epsilon 2^m}{c})SW_{AU_{\epsilon}}(A_{k}^{*}) \nonumber \\
&\hspace{-20pt} \geq (1-\frac{2\epsilon 2^m}{c}) \nonumber \\
&\hspace{-20pt} \quad \times\left(\max_{A_k=[a_{i_{1}},\ldots,a_{i_{k}}]}SW(A_k) - O\left(\frac{m\log{n}}{\epsilon n}\right)\right) \nonumber\\
&\hspace{-20pt}\geq \max_{A_k=[a_{i_{1}},\ldots,a_{i_{k}}]}SW(A_k) - \frac{2\epsilon 2^m}{c}  - O\left(\frac{m\log{n}}{\epsilon n}\right), \nonumber
\end{flalign}
where the second inequality holds with high probability according to Theorem \ref{thm:refined}.

Now setting $\epsilon=O\left(\sqrt{\frac{c m\log{n}}{2^m n}}\right)$ completes the result.
\qed

Therefore for large enough $n$, while ensuring truthfulness, the performance of our method converges to the simple algorithm that we mentioned at the start of this section.

\section{Discussion}
In this paper, we introduce an approach that uses the Differential Privacy framework to maintain auction truthfulness in a combinatorial auction setting. It's worth mentioning that our algorithm is a simplified version of MLCA, with the query phase only in the initial stage. By using the Differential Privacy framework, we can facilitate the truthfulness of the auction asymptotically and non-asymptotically. However, several assumptions in our design limit it to specific applications. For instance, the time complexity is not applicable when n and m are large, as it becomes $O(n^{mk})$. To address these limitations, future work can be done.

\section{Conclusions}
In this paper, we propose a new approach to convert combinatorial auctions that utilize machine learning into truthful mechanisms. The main goal is to compel rational bidders to reveal their true valuation functions during the auction process. Our method guarantees the accuracy of information provided by bidders and ensures that if the final allocation of the auction has a high expected social welfare, our method also yields a correspondingly high social welfare expectation. This means that our approach achieves truthfulness while also maintaining the efficiency of the original design of the combinatorial auction.

\bibliographystyle{IEEEtran}
\bibliography{main}

\end{document}